\documentclass[twocolumn]{aastex631}

\usepackage{booktabs}
\usepackage{tabularx}

\received{28-Jan, 2025}
\revised{11-Apr, 2025}
\accepted{17-Apr, 2025}

\begin{document}

\title{An Unsupervised Machine Learning Approach to Identify Spectral Energy Distribution Outliers:
Application to the S-PLUS DR4 data \footnote{Released on March, 1st, 2021}}

\author[0000-0001-8741-8642]{F. Quispe-Huaynasi}
\affiliation{Observatório Nacional, Rio de Janeiro, RJ 20921-400, Brazil}

\author[0000-0001-7059-5116]{F. Roig}
\affiliation{Observatório Nacional, Rio de Janeiro, RJ 20921-400, Brazil}

\author[0000-0002-8504-6248]{N. Holanda}
\affiliation{Observatório Nacional, Rio de Janeiro, RJ 20921-400, Brazil}

\author[0000-0002-0134-2024]{V. Loaiza-Tacuri}
\affiliation{Observatório Nacional, Rio de Janeiro, RJ 20921-400, Brazil}
\affiliation{Departamento de Física, Universidade Federal de Sergipe, São Cristóvão, SE 49100-000, Brazil}

\author[0009-0003-6830-8044]{Romualdo Eleutério}
\affiliation{Observatório Nacional, Rio de Janeiro, RJ 20921-400, Brazil}

\author{C. B. Pereira}
\affiliation{Observatório Nacional, Rio de Janeiro, RJ 20921-400, Brazil}

\author[0000-0001-9205-2307]{S. Daflon}
\affiliation{Observatório Nacional, Rio de Janeiro, RJ 20921-400, Brazil}

\author[0000-0003-4479-1265]{V. M. Placco}
\affiliation{NSF NOIRLab, 950 N. Cherry Ave., Tucson, AZ 85719, USA}

\author[0000-0002-6211-7226]{R. Lopes de Oliveira}
\affiliation{Departamento de Física, Universidade Federal de Sergipe, São Cristóvão, SE 49100-000, Brazil}

\author[0000-0002-3182-3574]{F. Sestito}
\affiliation{Centre for Astrophysics Research, University of Hertfordshire, Hatfield, AL10 9AB, UK}

\author[0000-0003-3537-4849]{P. K. Humire}
\affiliation{Instituto de Astronomia, Geofísica e Ciências Atmosféricas, Universidade de São Paulo, São Paulo, SP 05508-900, Brazil}

\author[0000-0001-5740-2914]{M. Borges Fernandes}
\affiliation{Observatório Nacional, Rio de Janeiro, RJ 20921-400, Brazil}

\author[0009-0007-8005-4541]{A. Kanaan}
\affiliation{Departamento de Física, Universidade Federal de Santa Catarina, Florianópolis, SC 88040-900, Brazil}

\author[0000-0002-7736-4297]{C. Mendes de Oliveira}
\affiliation{Instituto de Astronomia, Geofísica e Ciências Atmosféricas, Universidade de São Paulo, São Paulo, SP 05508-900, Brazil}

\author[0000-0002-0138-1365]{T. Ribeiro}
\affiliation{Rubin Observatory Project Office, Tucson, AZ 85719, USA}

\author[0000-0002-4064-7234]{W. Schoenell}
\affiliation{GMTO Corporation, Pasadena, CA 91107, USA}

%\collaboration{20}{(AAS Journals Data Editors)}

\begin{abstract}

Identification of specific stellar populations using photometry for spectroscopic follow-up is a first step to confirm and better understand their nature. In this context, we present an unsupervised machine learning approach to identify candidates for spectroscopic follow-up using data from the Southern Photometric Local Universe Survey (S-PLUS). First, using an anomaly detection technique based on an autoencoder model, we select a large sample of objects ($\sim 19,000$) whose Spectral Energy Distribution (SED) is not well reconstructed by the model after training it on a well-behaved star sample. Then, we apply the t-distributed Stochastic Neighbor Embedding (t-SNE) algorithm to the 66 color measurements from S-PLUS, complemented by information from the SIMBAD database, to identify stellar populations. Our analysis reveals 69 carbon-rich star candidates that, based on their spatial and kinematic characteristics, may belong to the CH or Carbon-Enhanced Metal-Poor (CEMP) categories. Among these chemically peculiar candidates, we identify four as likely carbon dwarf stars. We show that it is feasible to identify three primary white dwarf (WD) populations: WDs with hydrogen-dominated atmospheres (DA), WDs with neutral helium-dominated atmospheres (DB), and the WDs main sequence binaries (WD + MS). Furthermore, by using eROSITA X-ray data, we also highlight the identification of candidates for very active low-mass stars. Finally, we identified a large number of binary systems using the autoencoder model, but did not observe a clear association between the overdensities in the t-SNE map and their orbital properties.

%%%%%%%%%%%%%%%%%%%%%
%Studying peculiar stellar populations is essential for understanding the processes that generate their unusual properties, which in turn can help constrain models of stellar evolution. Consequently, identifying these atypical sources becomes highly valuable. In this work, we employ anomaly detection techniques using an autoencoder architecture to select candidates for anomalous sources in data provided by the Southern Photometric Local Universe Survey (S-PLUS). We then apply the t-distributed Stochastic Neighbor Embedding (t-SNE) algorithm to the 66 color measurements from S-PLUS, complemented by information from the SIMBAD database, to identify peculiar objects. Our analysis reveals 75 carbon stars candidates that, based on their spatial and kinematic characteristics, may belong to the CH or CEMP populations. Among these chemically peculiar candidates, we identify four as likely carbon dwarf stars. We report 5 candidate cataclysmic variables, two of which have observations in the eROSITA X-ray catalog, supporting their classification. We show that it is feasible to identify three primary white dwarf populations: DA, DB, and the white dwarf-main sequence. Furthermore, by utilizing X-ray information, we also highlight the identification of very active low-mass stars.
%%%%%%%%%%%%%%%%%%%%%

\end{abstract}

%% Keywords should appear after the \end{abstract} command. 
%% The AAS Journals now uses Unified Astronomy Thesaurus concepts:
%% https://astrothesaurus.org
%% You will be asked to selected these concepts during the submission process
%% but this old "keyword" functionality is maintained in case authors want
%% to include these concepts in their preprints.

%%\keywords{Peculiar stars (251) --- Carbon stars (1736) --- White dwarfs (1868) --- Low mass stars (804) --- Binary stars ()}

%% From the front matter, we move on to the body of the paper.
%% Sections are demarcated by \section and \subsection, respectively.
%% Observe the use of the LaTeX \label
%% command after the \subsection to give a symbolic KEY to the
%% subsection for cross-referencing in a \ref command.
%% You can use LaTeX's \ref and \label commands to keep track of
%% cross-references to sections, equations, tables, and figures.
%% That way, if you change the order of any elements, LaTeX will
%% automatically renumber them.
%%
%% We recommend that authors also use the natbib \citep
%% and \citet commands to identify citations.  The citations are
%% tied to the reference list via symbolic KEYs. The KEY corresponds
%% to the KEY in the \bibitem in the reference list below. 

\section{Introduction}\label{sec:intro}

Over the past two decades, spectroscopic surveys at low, medium and {high resolution, have contributed to the characterization of specific stellar populations and the discovery of peculiar objects. Examples are the Large Sky Area Multi-Object Fiber Spectroscopic Telescope survey \citep[LAMOST;][]{Cui2012}, the RAdial Velocity Experiment \citep[RAVE;][]{2006AJ....132.1645S}, the Apache Point Observatory Galactic Evolution Experiment \citep[APOGEE;][]{Majewski2017}, the GALactic Archaeology with HERMES survey \citep[GALAH;][]{2015MNRAS.449.2604D}, among others.
More recently, studies using Gaia DR3 low-resolution spectra ($R \approx 50$) from the BP and RP photometers (Gaia BP/RP), along with machine learning techniques, have reported a large sample of candidate stars, including carbon-enhanced metal-poor stars, metal-poor stars, and white dwarfs \citep{2023MNRAS.523.4049L, 2024MNRAS.52710937Y, 2024ApJ...970..181K}.
However, both currently and in the past, the identification of distinct stellar populations and candidates to peculiar objects for later spectroscopic follow-up was primarily facilitated by multi-band photometric surveys such as the Sloan Digital Sky Survey \citep[SDSS;][]{2000AJ....120.1579Y}\footnote{which also had low-resolution spectroscopy} and the Panoramic Survey Telescope and Rapid Response System \citep[Pan-STARRS;][]{2016arXiv161205560C}, among others. In addition to enabling the identification and classification of stellar populations using color-magnitude or color-color diagrams \citep{2000AJ....120.2615F, 2003ApJ...586..195H, 2002ASPC..261..297S, 2004A&A...418...77M}, standard photometric systems can provide information on stellar parameters \citep{1998ApJS..119..121L, 2006A&A...450..735M}. Furthermore, combined with infrared and/or ultraviolet data from photometric surveys like the Two-Million All Sky Survey \citep[2MASS;][]{2006AJ....131.1163S}, WISE \citep{2010AJ....140.1868W}, and GALEX \citep{2005ApJ...619L...1M}, the precision of this information can be improved \citep{2014AstBu..69..160S, 2016A&A...595A.129A}.

On the other hand, photometric surveys like Pristine \citep{2017MNRAS.471.2587S}, SkyMapper \citep{2019MNRAS.489.5900D}, the Javalambre Photometric Local Universe Survey \citep[J-PLUS,][]{2019A&A...622A.176C}, the Southern Photometric Local Universe Survey \citep[S-PLUS,][]{2019MNRAS.489..241M}\footnote{\url{https://www.splus.iag.usp.br/}}, and the Javalambre Physics of the Accelerating Universe Astrophysical Survey \citep[J-PAS,][]{2014arXiv1403.5237B}, which combine broad-band and narrow-band filters, have shown greater potential for the identification of stellar populations and the determination of stellar parameters with very good accuracy. For instance, the H filter (J0660) and the Ca II triplet (J0861) in the S-PLUS, J-PLUS, and JPAS surveys can capture features that indicate stellar activity and chemical composition. Moreover, the J0660 filter facilitates the identification of H-emitting low-mass stars, which are often undetectable with broad-band filters alone.

Recently, using machine learning techniques and photometric data from these surveys, it has become possible not only to determine stellar parameters but also to obtain chemical information for millions of stars \citep{2021ApJ...912..147W, 2024arXiv240802171H, 2024arXiv241118748F}. Leveraging the potential of combining broad-band and narrow-band filters, this study aims to identify outlier spectral energy distributions (SEDs) derived from the S-PLUS photometric data, using an anomaly detection technique with an autoencoder architecture. Subsequently, by applying a t-SNE algorithm to the colors derived from the 12 magnitudes provided by S-PLUS, alongside data from the SIMBAD catalog, we identify candidate M-type stars, carbon-rich stars, binary stars, white dwarfs, and other stellar populations within the anomalous sample.

This article is structured as follows: Section \ref{sec:splus_data} provides a brief overview of the S-PLUS data used in this study. Section \ref{sec:three} outlines the process of selecting a large sample of stars that deviate from the expected behavior of well-behaved stars, employing anomaly detection technique with autoencoder. Section \ref{sec:sect-fourth} details the use of the t-SNE dimensionality reduction to identify stellar subpopulations within the sample selected by the autoencoder. The characteristics of some of the identified populations are discussed in Section \ref{sec:pec_stars}. Finally, Section \ref{sect:disc_conc} presents the discussion and conclusions.

\section{S-PLUS data} \label{sec:splus_data}

The Southern Photometric Local Universe Survey is a photometric survey that observes the southern hemisphere with a dedicated 0.8 m robotic telescope hosted at the Cerro Tololo Inter-American Observatory, Chile. The telescope is equipped with a camera (T80Cam-S) of $9.2\mathrm{k} \times 9.2\mathrm{k}$ pixels, that provides a $2\deg^2$ field of view (FoV) with a pixel scale of 0.55~$\mathrm{arsec\,pix}^{-1}$. The observational strategy, image reduction, and main scientific goals of S-PLUS are described in \citet{2019MNRAS.489..241M}. S-PLUS combines seven narrow-band filters (J0378, J0395, J0410, J0430, J0515, J0660, and J0861) and five broad-band filters (u, g, r, i, and z) in the optical range ($\sim 3500$-10000 \AA). The g, r, i, and z band filters are similar to the corresponding SDSS filters, while the u band filter is the Javalambre u filter, which is an improved version os the SDSS u filter in terms of transmission. The characteristics and information about the filters are shown in Table \ref{tab:splus_filters} \citep{2019MNRAS.489..241M}. 

The S-PLUS Data Release Four (S-PLUS-DR4) includes 1,629 fields, covering about 3,000 $\deg^2$ of the southern sky, all reduced and calibrated in all the survey bands. Since this study exclusively uses data from the S-PLUS DR4 \citep{2024A&A...689A.249H} and from the Gaia Data Release 3 \citep[Gaia DR3,][]{2023A&A...674A...1G}, hereafter we will refer to them simply as S-PLUS and Gaia, respectively.

\begin{table}
\centering
\caption{S-PLUS photometric bands}
\label{tab:splus_filters}
\begin{tabularx}{\columnwidth}{lccc}
\hline 
Filter & Central wavelength & $\Delta \lambda$ & Spectral features \\
       &   \AA  &  \AA  &                                      \\
\hline
uJAVA     &   3563  & 352  & Balmer-dicontinuity region                       \\
J0378 &   3770  & 151  & [OII]                                     \\
J0395 &   3940  & 103  & Ca\,H + K                                 \\
J0410 &   4094  & 201  & $\mathrm{H}_{\delta}$                     \\
J0430 &   4292  & 201  & G band                                    \\
gSDSS     &   4751  & 1545 & SDSS-like g                               \\
J0515 &   5133  & 207  &  Mg \textit{b} triplet                    \\
rSDSS     &   6258  & 1465 &  SDSS-like r                              \\
J0660 &   6614  & 147  &  $\mathrm{H}_{\alpha}$                    \\
iSDSS     &   7690  & 1506 &  SDSS-like i                              \\
J0861 &   8611  & 408  &  Ca triplet                               \\
zSDSS     &   8831  & 1182 &  SDSS-like z                              \\
\hline
\end{tabularx}
\end{table}

\section{Anomaly detection and peculiar candidates selection}\label{sec:three}

We used an autoencoder architecture \citep{1986Natur.323..533R, 2006Sci...313..504H} to identify stars that exhibit unique spectral energy distributions (SED) within the S-PLUS dataset. There are different techniques for performing this task, however, here we use the simplest form of autoencoder (the so-called vanilla autoencoder), because it is widely used in the literature, it is easy to implement, and it is more intuitive. Classified as an unsupervised neural network, this architecture operates by first compressing the input data into a lower-dimensional representation\footnote{This low-dimensional representation is refereed to as code or latent space representation in the literature.} (the encoder), followed by the approximate reconstruction of the original data based on the lower representation (the decoder). The goal of training an autoencoder is to minimize the reconstruction error or loss function, which is the measure of the difference between the original input and its reconstructed output. The encoder and decoder both consist of fully connected traditional neural networks. 
Due to their characteristics, autoencoder architectures and extensions have been widely used for dimensionality reduction tasks \citep{HintonSalakhutdinov2006b,2015MNRAS.452..158Y, 2020AJ....160...45P} and as generative models \citep{8354080, NAZABAL2020107501}. Recently, a novel architecture called the scatter variational autoencoder was developed, which learns to generate a Gaia BP/RP spectra and estimate intrinsic scatter \citep{2025ApJ...979....5L}. While autoencoders have multiple applications across various domains \citep{Bank2023, Berahmand2024, Mienye2025}, our focus in this study is on their ability for the detection of anomalies in the SED. In this context, a similar approach using autoencoder has been applied to SDSS spectra by \citet{2019BSRSL..88..174S}, to identify potentially anomalous objects.

In general, the goal of anomaly detection is to find data instances that do not follow the expected pattern, or that deviate significantly from the majority of the data instances, representing a small set of the total sample \citep{10.1145/3439950}. In particular, our aim here is apply an autoencoder architecture to identify stars with atypical SEDs. The idea is to train the autoencoder with a sample of normal SEDs (majority class) or non-anomalous SEDs to learn to recognize the typical patterns of this normal data sample. After the training, it is expected that the reconstruction error values (anomaly score) will be lower for the normal data compared to the values for the anomalous data.

\subsection{Data training}

The training sample was obtained of an S-PLUS Value Added Catalog, where the sources were classified as stars (CLASS = 1) according to the QSO/star/galaxy classification described in \cite{2021MNRAS.507.5847N}. Since the goal is to select well-behaved sources, with reliable photometric and astrometric parameters, several quality cuts have been applied. Considering the photometry from S-PLUS, only objects with magnitudes lower than 19 and magnitude errors below 0.15 in all the 12 filters have been selected. Furthermore, we performed a cross-match between S-PLUS and Gaia and applied the following quality cuts: (i) $\mathtt{phot\_bp\_rp\_excess\_factor < 1.3}$ \citep[for normal stars, which are neither close binaries, nor situated in crowded regions, this factor should be close to 1,][]{2018A&A...616A...4E}; (ii) sources with parallax greater than zero ($\mathtt{\varpi > 0}$); (iii) $|\mathtt{astrometric\_gof\_al}| < 3$ and $\mathtt{ruwe < 1.2}$ (where $\mathtt{ruwe}$ is the Re-normalised Unit Weight Error, to ensure a good astrometric solution and avoid binary systems); (iv) $\mathtt{number\_visibility\_periods} > 15$ (as recommended by Gaia, sources with visibility periods $>9$ should be preferable); (v) $\mathtt{in\_qso\_candidates = 0}$ and $\mathtt{in\_galaxy\_candidates = 0}$ (to remove probably quasar and galaxy candidates, respectively); (vi) $\mathtt{phot\_variable\_flag \neq VARIABLE}$ (which excludes sources classified as variable); (vii) $\mathtt{non\_single\_star = 0}$ (that filters out sources identified as non-single stars, like astrometric binaries, spectroscopic binaries, or eclipsing binaries); and (viii) $\mathtt{duplicated\_source = 0}$ (to eliminate duplicated sources). 

In addition to the above criteria, we consider sources with $\mathtt{fidelity\_v2} > 0.5$ and $\mathtt{norm\_dg} < -3$ flags,\footnote{These flags were obtained by training a neural network, and were originally included in the external table \texttt{gedr3spur.main}, hosted at the German Astrophysical Virtual Observatory (GAVO), which is currently available in the table \texttt{external.gaiaedr3\_spurious} at \url{https://gea.esac.esa.int/archive/}.} as described in \citet{2022MNRAS.510.2597R} catalog. These cuts are used to select sources with reliable astrometric solutions and to exclude sources with potential color contamination from nearby objects, respectively.

After applying this sequence of quality cuts, the initial sample of approximately 3.5 million sources in S-PLUS is reduced to about 1.5 million objects, which will constitute our training data set. To characterize the training sample, we cross-matched it with the SIMBAD database, identifying $\sim 33,000$ objects, of which $\sim 20,000$ have information on stellar parameters: effective temperature ($T_{\mathrm{eff}}$), surface gravity ($\log g$), and metallicity ([Fe/H]). 
Figure \ref{fig:kiel_diagram} presents the color-magnitude diagram of the entire training data set, and the Kiel diagram ($\log g$ vs. $T_{\mathrm{eff}}$) of the subset with known parameters. This indicates that the dominant population of stars in the training sample has $T_{\mathrm{eff}}$ ranging from 4000 K to 7000 K, $\log g$ between 1.2 and 4.8 dex, and [Fe/H] spanning from -2 to +0.5 dex.

\begin{figure*}[!ht]
\centering
\includegraphics[width=\linewidth]{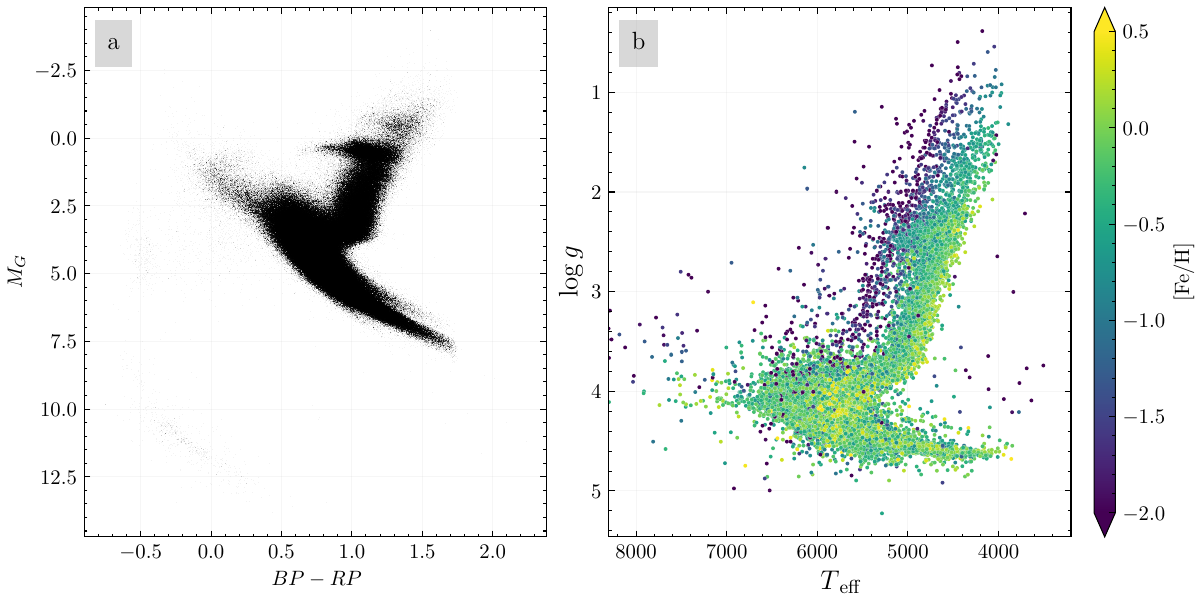}
\caption{(a) Color-magnitude diagram for the training sample ($\sim 1.5$ million stars). (b) Kiel diagram for objects in common between the training sample and SIMBAD, colored by metallicity ($\sim 20,000$ stars). The stellar parameters are those given in SIMBAD, and come from a diversity of independent works.}
\label{fig:kiel_diagram}
\end{figure*}

\subsubsection{Autoencoder architecture}
Once we have obtained our training sample, the next step is to define the autoencoder architecture that will be trained to select objects that do not follow the expected pattern for most S-PLUS stars. The autoencoder architecture used in this work is shown in Figure~\ref{fig:autoencoder_architecture} and was implemented through \texttt{keras} \citep{chollet2015keras}. Hyperparameters, including the number of hidden layers for both the encoder and decoder, neuron per layer, and the dimension of the latent space, were determined with automatic hyperparameters tuning using the \texttt{RandomSearch} algorithm implemented in \texttt{KerasTuner} \citep{omalley2019kerastuner}. The final architecture of the model, after the hyperparameters tuning, consists of two hidden layers with 8 nodes each for both the encoder and the decoder layers, and a latent space representation of dimension 7. During the training process, we use the Adam optimizer \citep{2014arXiv1412.6980K}, with learning rate defined by default in Keras ($\eta = 0.001$), apply the ReLU activation function, and employ the Mean Squared Error (MSE) as the objective function. 

\begin{figure*}[!ht]
\centering
\includegraphics[width=\linewidth]{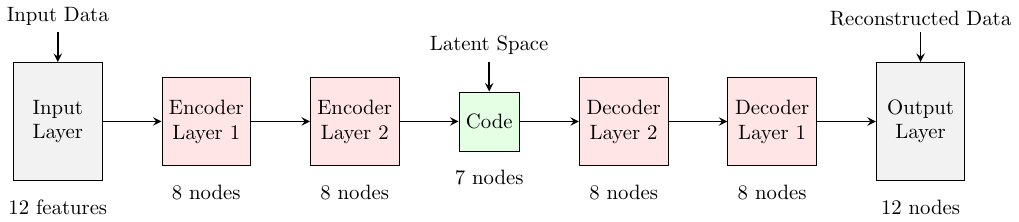}
\caption{Schematic representation of the autoencoder architecture after hyperparameter tuning. 
\label{fig:autoencoder_architecture}}
\end{figure*}

\subsection{Autoencoder applied to the S-PLUS data}

Giving as input for the autoencoder the 12 magnitudes from the S-PLUS filters, previously scaled between 0 and 1 using the MinMaxScaler() function provided by \texttt{Scikit-learn} \citep{scikit-learn}, we detected approximately $19,000$ stars (anomalous sources) that surpassed 3 standard deviations from the mean global reconstruction error found during the training phase. The left panel in Figure~\ref{fig:input-output_AE}, illustrates the input and output in blue and orange, respectively, for a source classified as normal (Gaia DR3 5659571422048804224). In contrast, the right panel displays the input and output for a source (Gaia DR3 2329317895999827840) identified as an anomalous star. It is evident that the autoencoder fails to reconstruct the input SED for the anomalous source. It is worth noting that the photometric errors of the input data are not taken into account in the comparison between the input and the output data. This is unnecessary because the photometric errors are, in general, much smaller than the threshold applied to the reconstruction error to classified the anomalous sources (see Sect.~\ref{sect:disc_conc}).

\begin{figure*}[!ht]
\centering
\includegraphics[width=\textwidth]{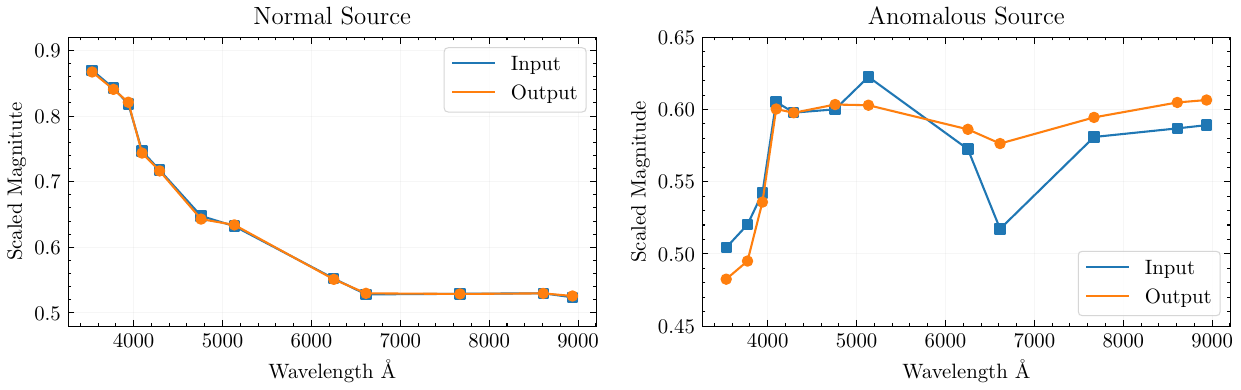}
\caption{Comparisons between input spectral energy distribution (SED), in blue, and the reconstructed SED, in orange, for a normal source (left panel) and for an anomalous source (right panel).}
\label{fig:input-output_AE}
\end{figure*}

\section{\lowercase{t}-SNE}\label{sec:sect-fourth}

To segregate stellar populations within the sample of objects identified as anomalous by the autoencoder, we applied the t-distributed Stochastic Neighbor Embedding algorithm \citep[t-SNE,][]{vanDerMaaten2008}, which is an extension of the Stochastic Neighbor Embedding algorithm \citep[SNE,][]{NIPS2002_6150ccc6}. t-SNE is a non-parametric unsupervised machine learning technique for dimensional reduction used for visualization, that has been applied in different areas of science, including astronomy \citep[among others]{2017ApJS..228...24T, 2018A&A...619A.125A, 2019MNRAS.483.3196C}. The main idea of this non-linear algorithm is to place objects described by a high-dimensional vector (features) into a low-dimensional space (map), preserving the local neighbor property defined in the high-dimensional space through a probabilistic approach. In this case, the similarity between points in the high dimensional space is described by a joint probability $p_{ij}$ given by
\begin{equation}
p_{ij} = \frac{p_{j|i} + p_{i|j}}{2n},
\end{equation}
where $n$ is the number of data points and
\begin{equation}
p_{j|i} = \frac{\exp\left(-|| x_i - x_j ||^2 / 2\sigma_i^2\right)}{\sum\limits_{k \neq i} \exp\left(-|| x_i - x_k ||^2 / 2\sigma_i^2\right)}
\end{equation}
is interpreted as the probability that the points $x_j$ are neighbors of the points $x_i$ based on the probability density of a Gaussian distribution with variance $\sigma_i$, centered at $x_i$. The value of $\sigma_i$ for each $x_i$ is determined internally by t-SNE through a binary search algorithm based on the user-defined perplexity parameter $(\mu)$ given by:
\begin{equation}
\mu = 2^{-\sum_j p_{j|i} \log_2 p_{j|i}}
\end{equation}
which can be interpreted as the effective neighbors around $x_i$. This procedure allows the algorithm to set smaller $\sigma_i$ values for dense regions and larger $\sigma_i$ values for more sparse regions.

Once $p_{ij}$ is defined, the next step is to define the joint probability $q_{ij}$ in the low-dimensional space using Student's t-distribution with one degree of freedom:
\begin{equation}
q_{ij} = \frac{(1 + \| y_i - y_j \|^2)^{-1}}{\sum_{k \neq l} (1 + \| y_k - y_l \|^2)^{-1}}.
\end{equation}

Considering that the initial positions of the points in the low-dimensional space are placed randomly or using Principal Component Analysis (PCA), t-SNE minimizes the Kullback-Leibler divergences (cost function) between $p_{ij}$ and $q_{ij}$, through the gradient-descent method, in order to preserve $p_{ij}$. 

\subsection{t-SNE applied to S-PLUS colors}

We apply the \texttt{Scikit-learn} implementation of the t-SNE in \textsc{python}\footnote{\url{https://scikit-learn.org/stable/modules/generated/sklearn.manifold.TSNE.html}} to a set of color indices, each one obtained as the difference between two magnitudes following the convention that the first magnitude always corresponds to the shorter-wavelength bandpass, and the second magnitude always corresponds to the longer-wavelength bandpass. By combining the 12 magnitudes provided by S-PLUS, we obtained a set of 66 color indices. These 66 indices are further normalized using the \texttt{StandardScaler()} function, also from \texttt{Scikit-learn}, to have zero mean and unit variance, and these constitute the input for t-SNE.
This analysis is performed on the sample of stars identified by the autoencoder as anomalous objects. Except for the perplexity and the number of iteration hyperparameters, which were set to 40 and 2000, respectively, all the other hyperparameters were left at their default values in the implementation.
Initially, we adopted a value of 30 for the perplexity, as recommended in \citet{vanDerMaaten2008}, and kept the other hyperparameters at their default settings. We then tested values of 25, 35, 40, and 45. Although there are some variations in the clumps' distribution and densities on the resulting 2D map, the changes are subtle for the tested values. As for the number of iterations, we tested 1000 (the default value), 2000, and 5000. Again, these values did not have a significant effect on the resulting map. This means that the results do not change significantly over a wide range of these two parameters. Choosing a perplexity value of 40, in particular, was mostly motivated because it yielded the highest number of carbon stars candidates.
Figure~\ref{fig:tsne_map} shows the t-SNE map, colored by the \texttt{ruwe} parameter and the absolute magnitude $M_G$. Over-densities appear and some regions are dominated by $\mathtt{ruwe} > 1.4$, suggesting the potential presence of binary systems in the sample. By running the t-SNE several times, we have verified that, despite some subtle variations in the map, the overall structure and densities remain mostly unchanged for the values adopted in this work.

To determine whether the clumps in the map correspond to previously classified objects, we crossmatched our data with the SIMBAD database, identifying approximately 8,000 cataloged sources. To identify the clumps, we use the primary classification in SIMBAD based on the \texttt{main\_type} label. To ensure a reliable classification and focus on specific populations, we excluded sources labeled as candidate, as well as those generically classified as `Star' (\texttt{main\_type$\neq$`Star'}). This refinement resulted in a final sample of 2,644 objects with specific classifications.

\begin{figure*}[!ht]
\centering
\includegraphics[width=\linewidth]{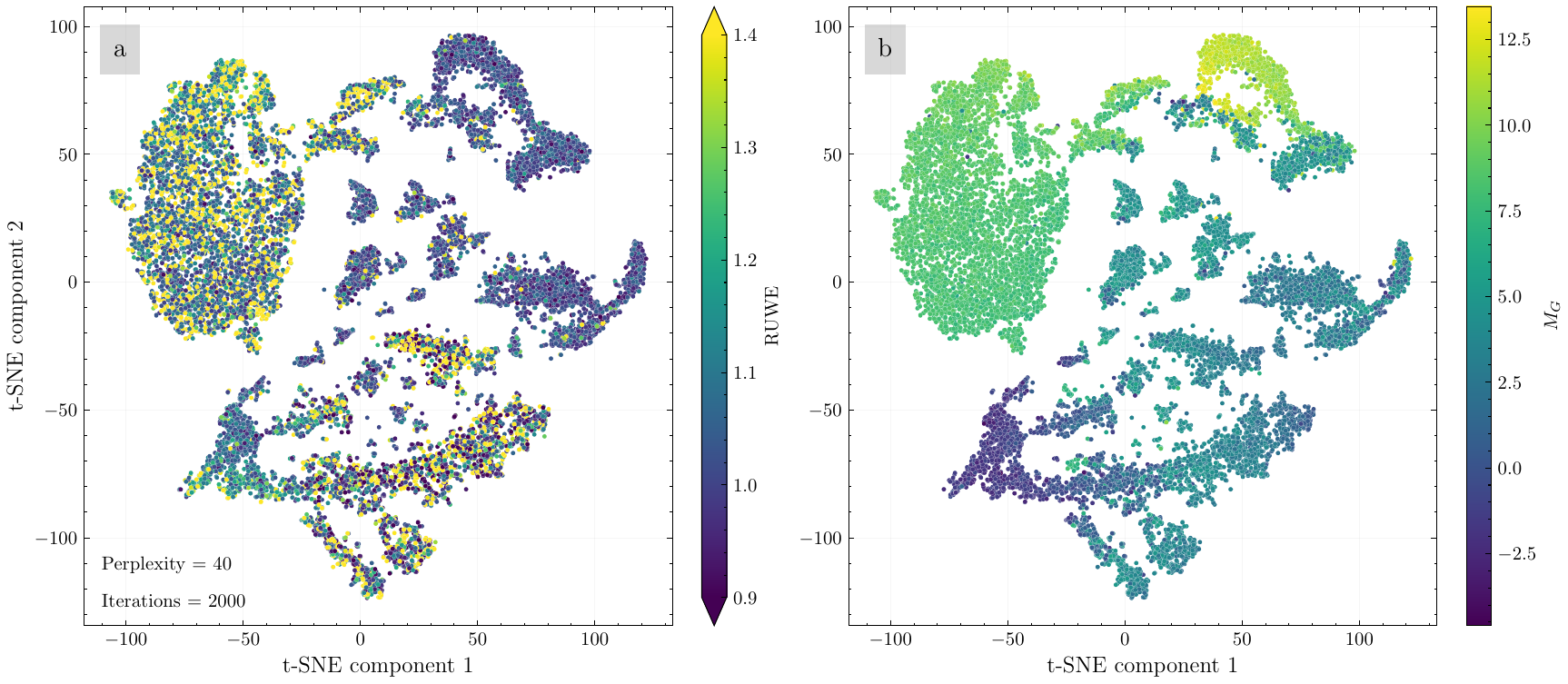}
\caption{t-SNE 2D projections. (a) Colored by Gaia \texttt{ruwe} parameter. (b) Colored by Gaia absolute magnitude. The maps contain all the sources identified using the autoencoder. Except for the perplexity and the number of iterations, which were set to 40 and 2000, respectively, all the other hyperparameters were kept at their default values in the \texttt{Scikit-learn} implementation of t-SNE. To compute the absolute magnitudes, we consider the distances provided by \citet{2021yCat.1352....0B}.}
\label{fig:tsne_map}
\end{figure*}

Figure~\ref{fig:tsne_projection_color} shows the t-SNE maps based on the specific populations identified in SIMBAD, with the labels reflecting their classifications and an asterisk (*) denoting the source as a star \footnote{Information about the classification in SIMBAD can be found at \url{http://simbad.u-strasbg.fr/guide/otypes.htx}}. The sources detected by the autoencoder are shown in gray, while the colored points represent the objects with information in SIMBAD. Some populations that are clearly grouped in these maps will be described in the next section. As a reference, Figure~\ref{fig:color-magnitude} presents the color-magnitude diagrams based on Gaia photometry for the sources shown in Figure~\ref{fig:tsne_projection_color} that have positive parallax and  photogeometric distances provided by \citet{2021yCat.1352....0B}. The x-axis in this figure is the $G_{BP} - G_{RP}$ color, and the y-axis is the absolute $\mathrm{G}$ magnitude ($M_G$). The latter has been determined using the apparent $G$ magnitude of Gaia, corrected by extinction using the SFD2D dust map \citep{Schlafly2011}, via the \texttt{dustmaps} \textsc{python} package \citep{Green2018}, along with the photogeometric distances. The extinction coefficients were calculated using the parameters listed in table 1 of \citet{2018A&A...616A..10G}, and applying a code by Koposov (private communication), which just implements the polynomials described in that paper.

\begin{figure*}[!ht]
\centering
\includegraphics[width=\textwidth]{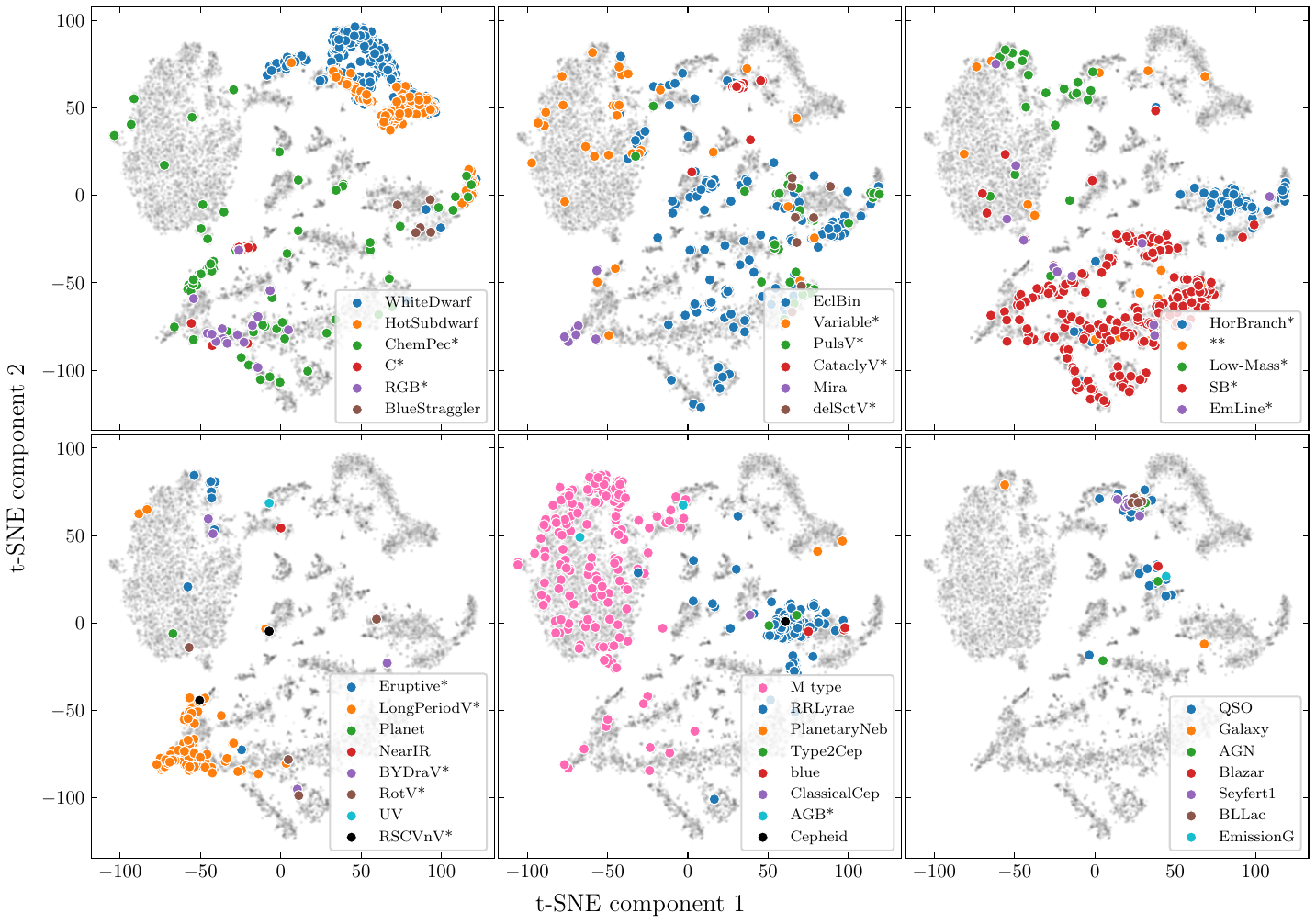}
\caption{t-SNE 2D maps: In all panels, the background in gray represents sources identified by the autoencoder. The colored points indicate sources that overlap with sources classified in the SIMBAD database. The ChemPec* label encompass a variety of chemically peculiar objects, including CEMPs, alpha2 CVn variables, R CrB variables, CH stars, barium stars, and dwarf carbon stars.}
\label{fig:tsne_projection_color}
\end{figure*}

\begin{figure*}[!ht]
\centering
\includegraphics[width=\textwidth]{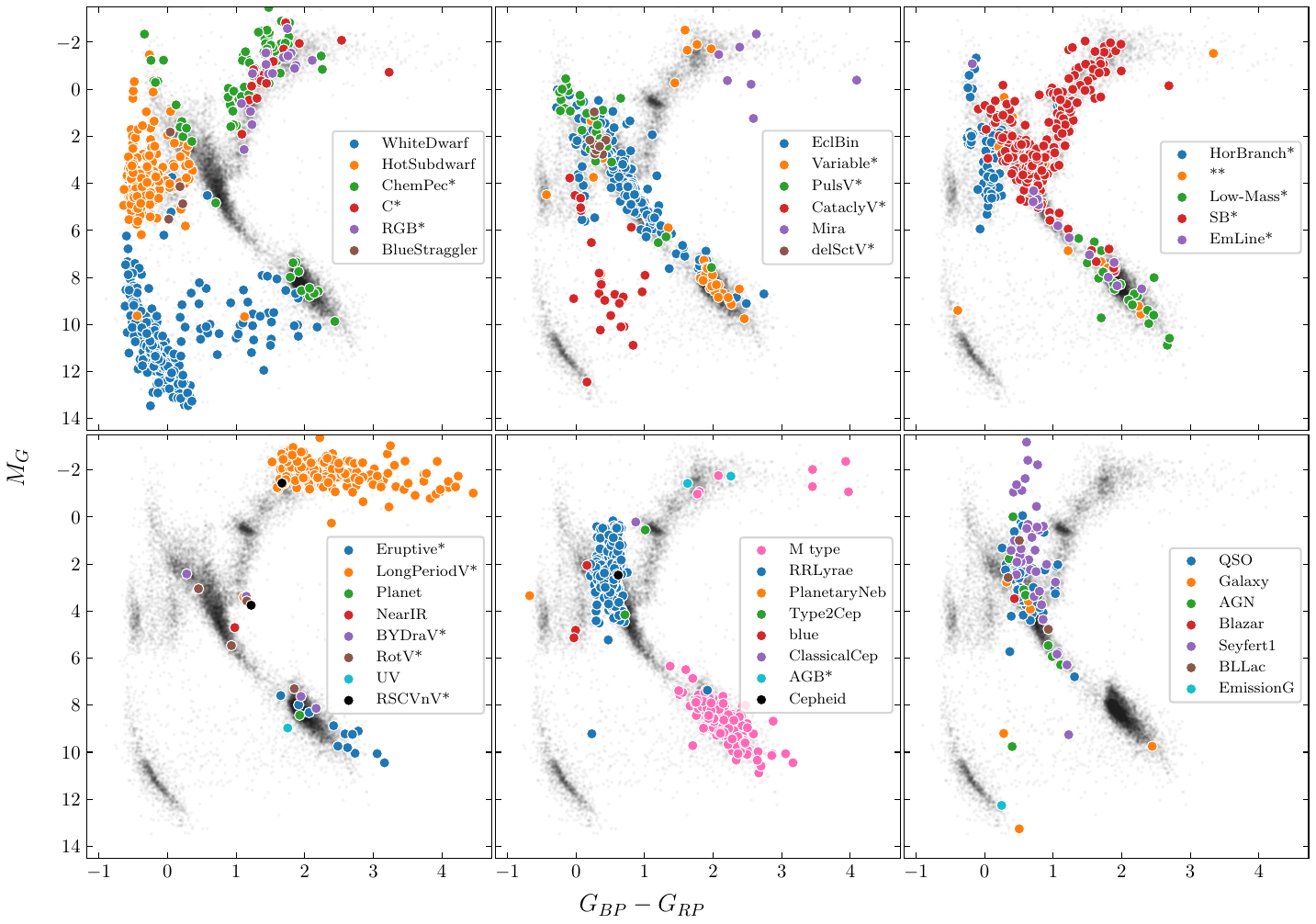}
\caption{Color-magnitude diagram. In all panels, the background in gray represents sources identified by the autoencoder. The colored points indicate sources that overlap with sources classified in the SIMBAD database. The ChemPec* label encompass a variety of chemically peculiar objects, including CEMPs, alpha2 CVn variables, R CrB variables, CH stars, barium stars, and dwarf carbon stars.} 
\label{fig:color-magnitude}
\end{figure*}

\section{Stellar populations identified using autoencoder and \lowercase{t}-SNE} \label{sec:pec_stars}

In this section, we focus on four distinct stellar classes that notably cluster in specific regions of the map after applying the t-SNE algorithm, The selection of these clusters has been done manually on the t-SNE map. Other populations, such as cataclysmic variables, or those that do not fit clearly into categories like the RR Lyrae stars, extragalactic sources, and others identified in this study, will be further analyzed in upcoming publications. Figure \ref{fig:sed_populations} shows, from top to bottom, the SEDs for sources selected as: a normal star, a carbon-rich star, a white dwarf star, an M-type star, and a single-line spectroscopic binary (SB1) system, respectively.  

\begin{figure}[!ht]
\centering
\includegraphics[width=\linewidth]{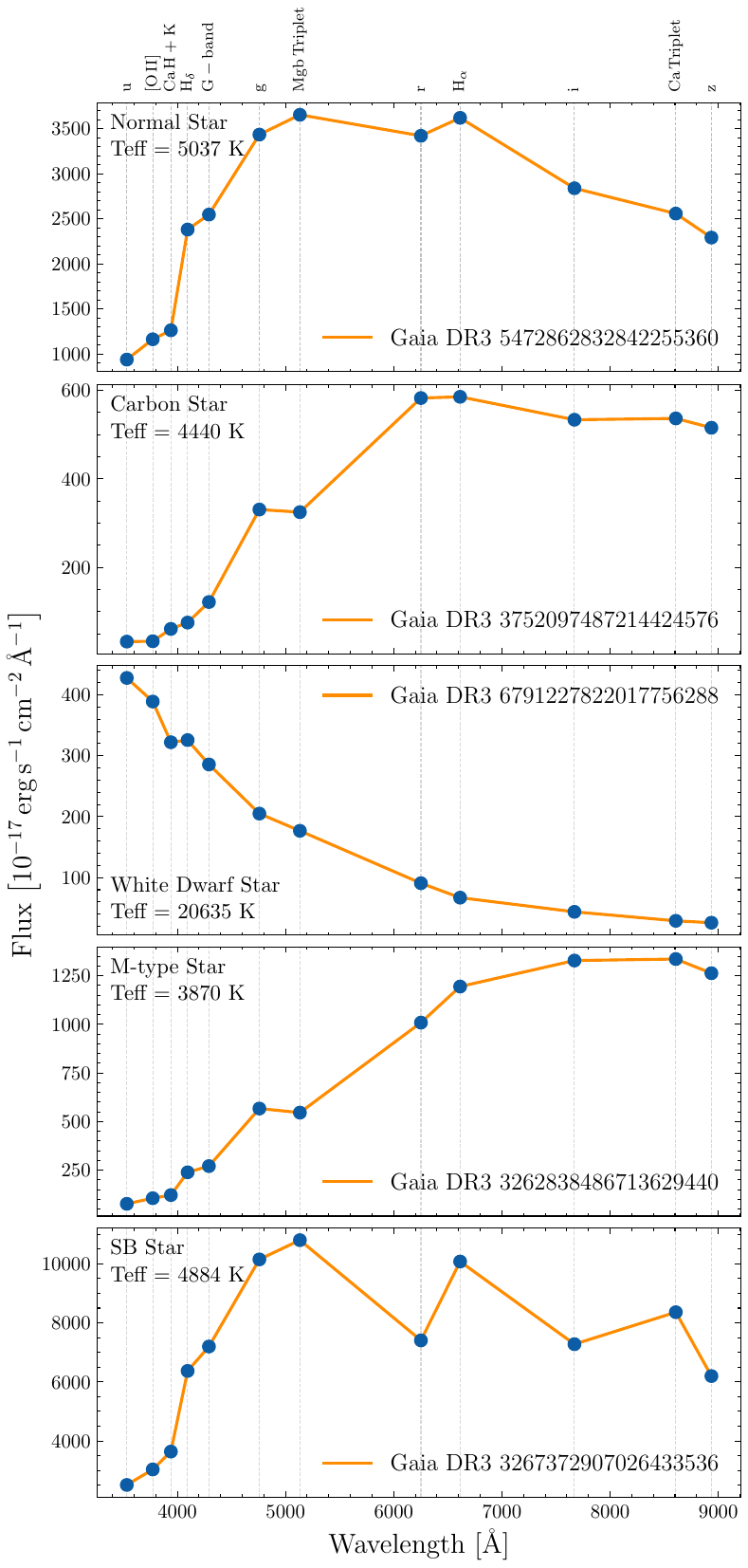}
\caption{Spectral energy distribution of the populations described in this work}
\label{fig:sed_populations}
\end{figure}

\subsection{Carbon Stars}

Carbon stars are defined by an overabundance of carbon relative to oxygen in their stellar atmospheres ($\mathrm{C/O > 1}$). Normal C stars, also known as C-N stars, are cool objects ($T_\mathrm{eff} <  3500$~K) in the thermally pulsing asymptotic giant branch (TP-AGB) phase. These stars become intrinsically carbon-rich as they enrich their atmospheres with carbon through third dredge-up events \citep{2005ARA&A..43..435H}. In addition to these intrinsic C stars, various types of C stars can form through extrinsic mechanisms, such as mass transfer in binary systems. Other scenarios, including mergers and anomalous He flashes that mix carbon to the surface \citep[e.g.,][]{zhang2020,zamora2009}, have also been proposed to explain the formation of different types of C stars.

Among these types, C-R stars (early and late) are warmer than C-N stars \citep[see][]{zamora2009}. Early-R stars have luminosities similar to red clump stars and show no s-process elements enhancement \citep{dominy1984,knapp2001}. Late-R stars are likely AGB stars, though not as evolved as C-N stars. C-J stars are characterized by very low isotopic carbon ratios, in some cases as low as $^{12}$C/$^{13}\mathrm{C}\approx 3.5$, and high lithium abundances, which are even higher than those of the early-R stars. Their luminosities are also consistent with the AGB phase \citep{abia2000}. CH stars, on the other hand, are found in a different metallicity regime ($-2\leq [\mathrm{Fe/H}] \leq -1$) and exhibit a very strong G band due to the CH molecule. They also have relatively high galactic velocities, typical of thick disk and halo stars. These population are expected to result from mass transfer from an AGB companion. At even lower metallicities, carbon-enhanced metal-poor stars \citep[CEMPs; see the complete classification by][]{beers2005} are thought to be the more metal-poor counterparts of CH stars. While most CEMP stars originate from mass transfer, some of them may have different formation pathways. Additionally, dwarf C stars (dC), which are found on the main sequence phase, can be identified using JHK color-color diagrams due to the weakness of their 2.3 $\mu$m CO bands compared to those of C giant stars \citep{wallerstein1998}.

With our method, we have identified 96 candidates C stars, with 27 of them previously reported in the literature \citep{stephenson1973,bothun1991,hayes2018,christlieb2001}. Assuming $M_{\rm G}\,>\,5.0$\,mag ($\log g > 4$) as a criterion, four of the newly identified objects are likely dC stars. 

To refine the classification of these candidates, we compare our sample in detail with the C stars studied by \citet{li2024}, who identified and classified 3546 C stars using the LAMOST DR\,7 data \citep{zhao2012}. Their classification relied on line indices, color-color diagrams, spatial distribution, and visual inspection of spectra. Figure\,\ref{fig:carbon_star_diagrams} presents a series of diagrams to differentiate various types of C stars and compare them with our candidates. Panel (a) shows a color-magnitude diagram that clearly identifies dwarf stars and distinguishes C-N stars, while the separation between CH and C-R stars is less distinct \citep[a subtle difference also noted by][in another set of C stars]{abia2022}. Panel (b) presents a kinematic diagram based on Gaia astrometry, including radial velocity measurements, which distinguishes between CH (thick disk) and C-R (thin disk) populations. In principle, Gaia provides radial velocities primarily for the brightest stars, meaning that we may be overlooking the fainter stars in our analysis. However, by applying a quality cut in parallax to the stars with radial velocity, we verified that the overall velocity pattern remained unaltered, implying that ignoring the fainter stars does not introduce an obvious bias toward any specific kinematic population. Panel (c) further supports that the C star candidates identified in this study are likely CH and CEMP stars with characteristics of thick disk and halo populations. By performing a cross-match with a recent catalog of candidate CEMP stars \citep{2023MNRAS.523.4049L}, we verified that 73 objects in our sample had already been classified as CEMP star candidates using BP/RP spectra from Gaia DR3, reinforcing their classification.

Figure\,\ref{fig:carbon-region} shows a zoom in of the top left panel in Figure\,\ref{fig:tsne_projection_color}, where these populations of carbon stars cluster in a specific region on the t-SNE map. This region overlaps with four stars classified as peculiar members of the CEMPs population and one RGB star. Table \ref{tab:carbon_sample} lists some of the candidate C stars; the complete sample is available in machine-readable form.

\begin{figure*}[!ht]
\centering
\includegraphics[width=\linewidth]{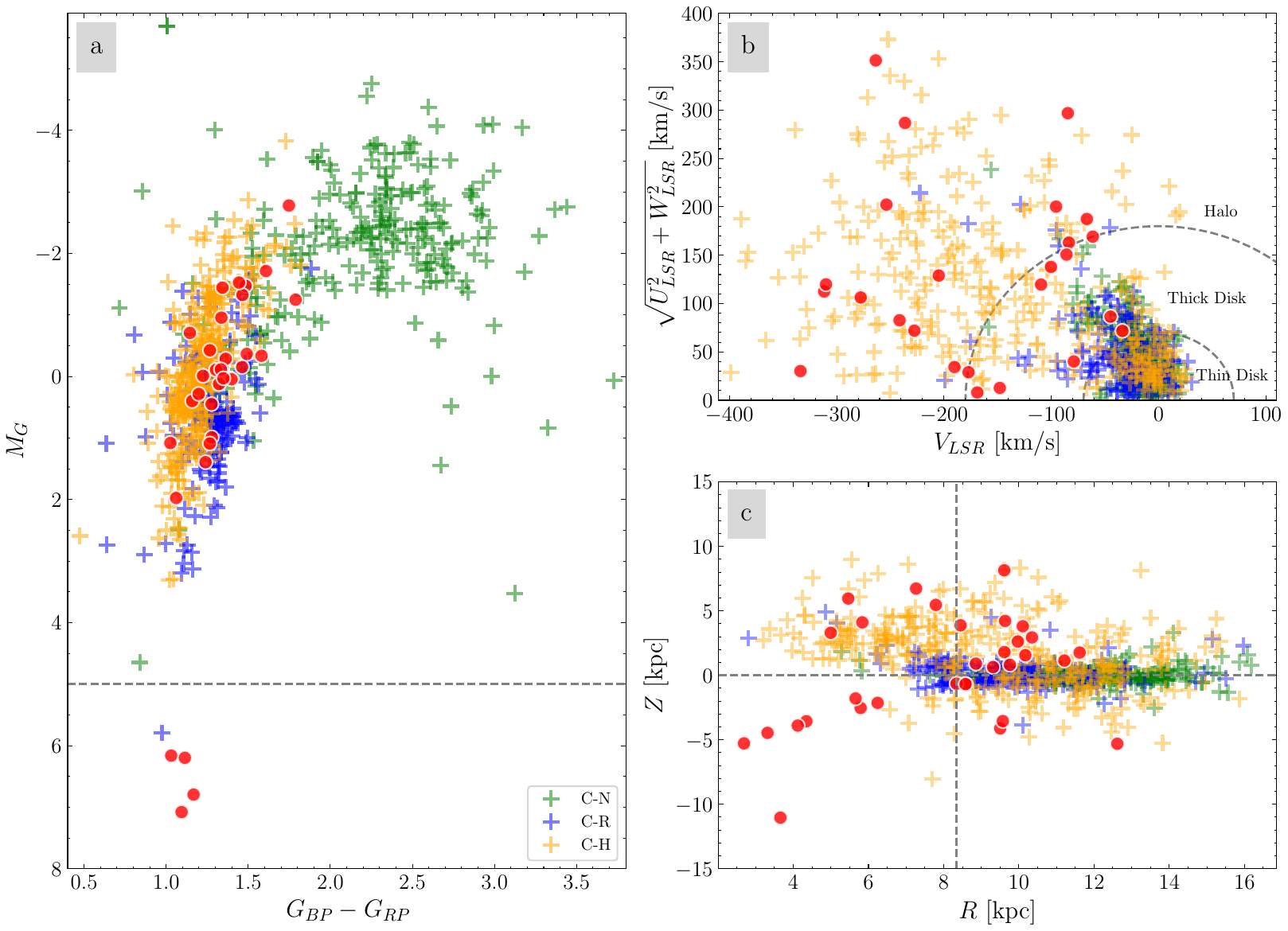}
\caption{Comparison of C star candidates (red circles) with the C stars analyzed by \citet{li2024}. Panel (a) presents a color-magnitude diagram with different types of C stars color-coded: green for C-N, orange for CH, and blue for C-R. A dashed line at $M_{\rm G}\,=\,5.0$\,mag marks the boundary for stars in the main sequence phase (dC). Panel (b) shows a Toomre diagram with dashed curves at $v_{t} = 180$ km\,s$^{-1}$ and 70 km\,s$^{-1}$, indicating the thick and thin disk regions, respectively. Panel (c) illustrates the distribution of objects by their Z coordinate as a function of the Galactocentric radius, R, assuming $R_{\odot} = 8.34$ kpc \citep{2014ApJ...783..130R}. Stars with a relative parallax error greater than 30\,\% were excluded from this analysis.}
\label{fig:carbon_star_diagrams}
\end{figure*}

\begin{figure}[!ht]
\centering
\includegraphics[width=\linewidth]{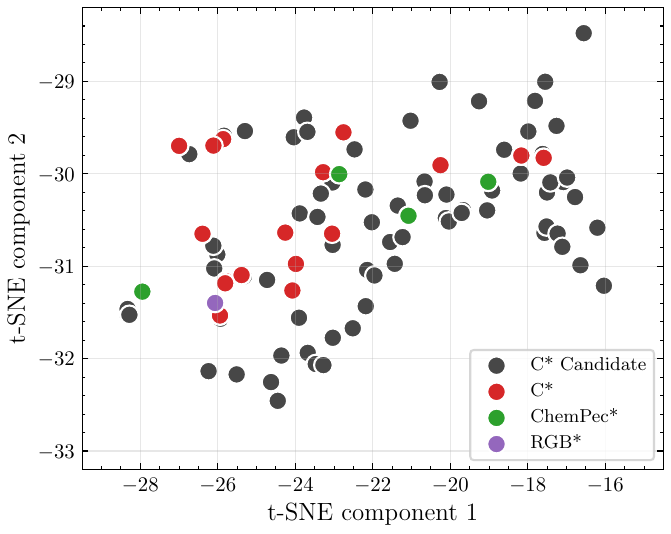}
\caption{Zooming in on the region of carbon-rich stars in the t-SNE map. The carbon-rich stars identified in this study are shown in gray. The red, green, and purple points indicate stars classified as carbon, peculiar, and RGB in SIMBAD, respectively.}
\label{fig:carbon-region}
\end{figure}

\begin{deluxetable*}{lccccccc}
\tablecaption{Astrometric parameters and G apparent magnitude from \textit{Gaia} DR3 for carbon-rich star candidates \label{tab:carbon_sample}}
\tablehead{
\colhead{\textit{Gaia} DR3 ID} & 
\colhead{$\alpha$} & 
\colhead{$\delta$} & 
\colhead{$\varpi$} & 
\colhead{$\mu_{\alpha}$} & 
\colhead{$\mu_{\delta}$} &
\colhead{$\mathrm{V_{rad} (Gaia)}$} &
\colhead{$\mathrm{G}$}  \\
\colhead{} & 
\colhead{$\mathrm{(deg \pm mas)}$} & 
\colhead{$\mathrm{(deg \pm mas)}$} & 
\colhead{$\mathrm{(mas)}$} & 
\colhead{$\mathrm{(mas\,yr^{-1})}$} &
\colhead{$\mathrm{(mas\,yr^{-1})}$} &
\colhead{$\mathrm{(km\,s^{-1})}$} &
\colhead{$\mathrm{(mag)}$} 
}
\startdata 
5658943497829715712 & 149.836 & -26.121 &  0.034 &  0.247 & -0.867 & 155.423 & 14.658 \\
5451431500088634240 & 162.409 & -31.364 &  0.128 & -2.096 & -0.931 &   NaN   & 16.307 \\
5458744764242802816 & 150.850 & -34.025 & -0.001 & -0.422 & -0.398 &   NaN   & 15.559 \\
5391795020192080896 & 161.673 & -41.746 &  0.040 & -4.519 &  1.264 &   NaN   & 14.456 \\
5391637446431500928 & 162.265 & -41.460 &  0.104 &  2.579 & -9.015 & 236.292 & 13.419 \\
\enddata
\tablecomments{This table is available in its entirety in machine-readable form.}
\end{deluxetable*}
\subsection{White dwarf stars}
White dwarfs (WD) mark the final evolutionary stage of stars with initial masses below $8M_{\odot}$. Classified as compact objects, they typically feature an extremely dense core ($\sim 10^{6}\,\mathrm{g\,cm^{-3}}$), primarily composed of carbon and oxygen, surrounded by a thin atmosphere \citep{ 2022PhR...988....1S, 2024arXiv240903941B}. Based on the chemical composition observed in their atmospheres, the main populations of white dwarfs can be classified as follows: DA (spectrum dominated by hydrogen lines), DB (spectrum dominated by neutral helium lines), DC (absence of spectral lines in their spectrum), DO (spectrum dominated by ionized helium), and DZ (spectrum with spectral lines from heavier elements). Since heavy elements are not expected in the atmospheres of WD, their presence indicates pollution with accreted planetary debris, and for this reason, such stars are known as polluted WD \citep{2014AREPS..42...45J}. 

%%%%%%%%%%%%%%%%%%%%%
%https://academic.oup.com/mnras/article/527/3/4515/7339778
%%%%%%%%%%%%%%%%%%%%%

Unlike other stellar populations, which occupy specific loci in the t-SNE map, the region of WDs partially overlaps with that of the hot sub-dwarfs (Figure~\ref{fig:tsne_projection_color}, top left panel). However, these populations are segregated in the color-magnitude diagram (Figure~\ref{fig:color-magnitude}, top left panel). In view of this, we add the $M_G$ magnitude from Gaia \citep[for those stars with available photogeometric distances from][]{2021yCat.1352....0B} to the corresponding S-PLUS colors and normalize the data using the \texttt{StandardScaler()} function; then used it as the input for the t-SNE. The resulting map is shown in Figure~\ref{fig:white_dwarf_tsne} (left panel), and we can see that the two populations are well segregated: the WD population is represented in blue and the hot sub-dwarfs population is represented in orange. 

Additionally, we can note that there are three regions of over-density in the WD population, labeled as 1, 2, and 3 in Figure~\ref{fig:white_dwarf_tsne}. To determine whether these over-densities correspond to specific sub-populations of WDs, we rely on sources that have spectral type information. The result is shown in Figure~\ref{fig:white_dwarf_tsne} (middle panel), where region 1 consists mostly of binary systems (WD + main sequence stars), region 2 mainly includes DB and DC type stars, and region 3 is primarily composed of DA type stars. In the right panel, we represent these populations in the color-magnitude diagram. Consistently with their classification, most of the binaries are located between the white dwarf sequence and the main sequence. In contrast, the DB and DC populations are not clearly separated from the DA stars in the color-magnitude diagram. We conclude that we can reliably apply the t-SNE to the S-PLUS colors to segregate the main sub-populations of white dwarfs. Nevertheless, we must mention that using color-color diagrams \citet{2022A&A...658A..79L} show segregation of the DA and DB populations in the J-PLUS data. Besides, although a clear segregation is not appreciable in the color-magnitude diagram, a large number of WD in \textit{Gaia} have revealed substructures associated with the WD sub-populations in the WD loci \citep{2018A&A...616A..10G}.

\begin{figure*}[!ht]
\centering
\includegraphics[width=\textwidth]{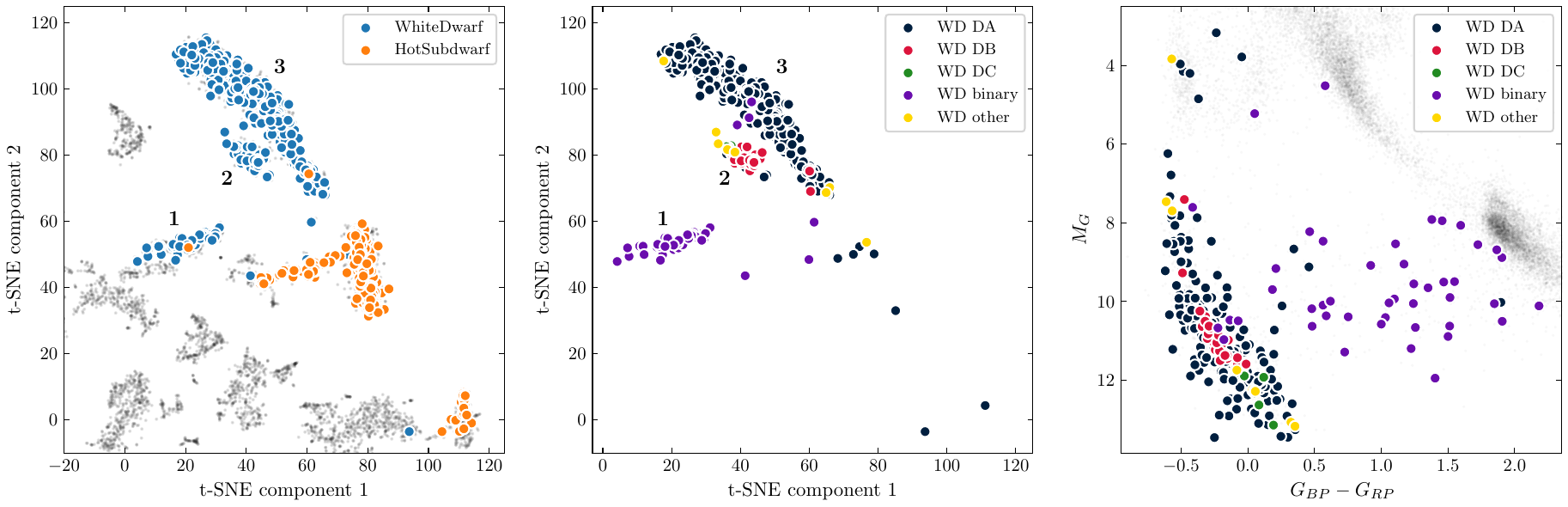}
\caption{White dwarf populations. The left panel shows the t-SNE map, illustrating the separation between WDs (blue dots) and hot subdwarf stars (orange dots) after using the $\mathrm{M_G}$ as an additional feature for the t-SNE input. The regions 1, 2, and 3 in this panel are over-densities on the t-SNE map. The middle panel shows the t-SNE map with spectral type information in SIMBAD for the same population of WD as in the left panel. In relation to the left panel, region 1 corresponds to binary systems (purple points), region 2 mainly corresponds to DB and DC stars (red and green points), and region 3 predominantly corresponds to DA stars (dark blue points). The right panel presents the color-magnitude diagram for WD. The locations of the WD populations in this diagram are consistent with their spectral classification.}
\label{fig:white_dwarf_tsne}
\end{figure*}

\subsection{Low mass stars}

One of the most numerous population ($\sim 5000$ stars) identified by the autoencoder is that of the low-mass stars ($M < 0.8 M_\odot$), located in the upper left part of the t-SNE map (Figure~\ref{fig:tsne_map}). According to data from SIMBAD (Figure~\ref{fig:tsne_projection_color}), this region contains low-mass stars, variable stars, and eruptive stars. Based on the spectral types of these populations, M-type stars are the predominant group in this area. Consistently with their spectral classification, these sources occupy the region of low-mass stars in the main sequence in the color-magnitude diagram (Figure~\ref{fig:color-magnitude}). From this sample, 416 stars have data from APOGEE. Considering that the effective temperatures are reliable in APOGEE for low-mass stars \citep{2021ApJ...917...11S, 2022ApJ...927..123S}, the reported temperatures are consistent with those expected for this population. 
%From this sample, 416 stars have effective temperature determined by APOGEE and fall within the range of $3400< T_{\rm eff}< 4100$~K, which is consistent with the expected values for low-mass stars.
Due to their low surface temperature, their spectra exhibit a complex structure, characterized by the presence of water molecules, methane, titanium oxide, and other compounds in their atmospheres. As illustrated in Figure~\ref{fig:sed_populations}, their SED differs significantly from that of the stars classified as normal in this study.

It is well known in the literature that some low-mass stars exhibit X-ray emission due to intense stellar activity \citep{2003ApJ...583..451M, 2004A&ARv..12...71G}. The origin of this activity is linked to the age, presence of strong magnetic fields and their rapid rotation, which contributes to the magnetic dynamo mechanism \citep{2017ApJ...834...85N}. Since X-ray data provide insights into coronal and chromospheric activity, as well as accretion processes in young low-mass stars, we present in Figure~\ref{fig:low_mass_sta_tsne} (upper panel) the sources that have X-ray information from the eRosita-DE Data Release 1 \citep{2024A&A...682A..34M}, highlighted in orange. The lower panel in the same figure shows the ``X-ray main sequence" diagram proposed by \cite{Rodriguez_2024}. This diagram distinguishes between accreting compact X-ray-emitting objects and stars that emit X-rays due to coronal activity, based on their optical color and X-ray-to-optical flux ratio $(F_\mathrm{X}/F_{\mathrm{opt}})$. Most of the sources are located in the active star region, below the empirical cutoff represented by the blue dashed line, as defined by \cite{Rodriguez_2024}. On the other hand, it is worth noting that in the t-SNE map, a large portion of the stars with X-ray information are clustered in the upper right part, slightly separated from the rest of the group. This result suggests that photometric data from S-PLUS can be used to identify candidates for highly active stars. This is not surprising, given that some of the S-PLUS filters are located over specific spectral lines such as Ca H + K, H$\delta$, H$\alpha$, and the Ca triplet, which are indicators of stellar activity \citep{2007A&A...469..309C}.

\begin{figure}[!ht]
\centering
\includegraphics[width=\linewidth]{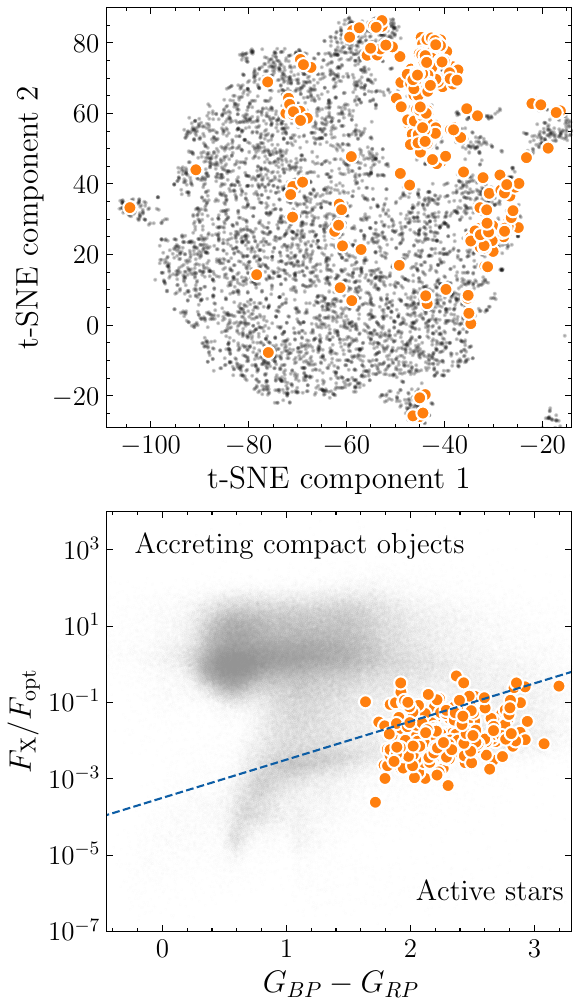}
\caption{The upper panel shows a zoomed-in view of the low-mass star region in the t-SNE map, with sources containing X-ray information highlighted in orange. The bottom panel displays the X-ray main-sequence diagram, where objects with X-ray data are positioned in the expected region for active stars.}
\label{fig:low_mass_sta_tsne}
\end{figure}

\subsection{Binary systems}\label{binary}

Because quality cuts, such as $\mathtt{astrometric\_gof\_al}$, $\mathtt{ruwe}$ and $\mathtt{non\_single\_star}$, have been applied to select the normal sources sample, binary systems are expected to be identified as anomalous sources. According to the \texttt{ruwe} parameter in Figure~\ref{fig:tsne_map}, the region occupied by the potential binary objects has a complex pattern in the t-SNE map. They are represented by the more yellowish colors, and are located mostly in the upper left and lower right regions. From the cross-match with SIMBAD (Figure~\ref{fig:tsne_projection_color}), most of the systems classified as spectroscopic binaries (SB) are located in the bottom part of the map, while the eclipsing binaries (EclBin) appear more dispersed throughout the map.  

Since binary systems are classified through their detection methods rather than by their physical properties, it is not reasonable to anticipate an over-density based on these classifications. Instead, to identify binary sub-populations within the anomalous sample, we search for stars that overlap with the \texttt{nss\_two\_body\_orbit} table from the non-single stars catalog provided by \textit{Gaia}, which includes the orbital parameters for astrometric, spectroscopic, and eclipsing binary systems \citep{2023A&A...674A..34G}. After the cross-match, we found $\sim400$ sources that are classified by solution type as \texttt{SB1}, \texttt{EclipsingBinary}, \texttt{Orbital}, or \texttt{AstroSpectroSB1}. Taking into account the information from the orbital parameters, it is not possible to observe a clear correlation with the over-densities in the map. This is likely due to the fact that multiple configurations of the orbital parameters can yield similar SEDs, although this issue would deserve a deeper analysis in the future.

\begin{figure}[!ht]
    \centering
    \includegraphics[width=\linewidth]{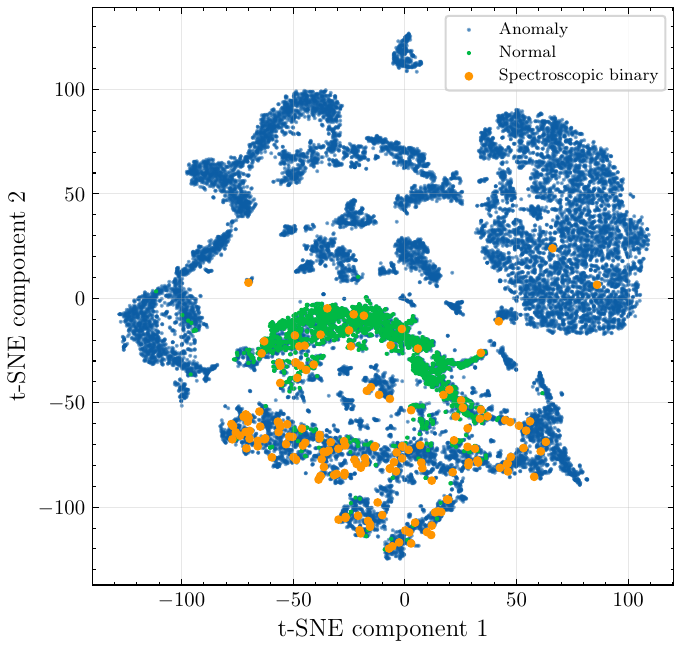}
    \caption{t-SNE 2D map constructed for a set that includes all the anomalous stars detected by the autoencoder (blue dots) and a subset of the normal stars used to train the autoencoder (green dots). The location of the anomalous stars identified as spectroscopic binaries in SIMBAD (orange dots) is plotted for reference.}
    \label{fig:anomaly-normal-tsne}
\end{figure}

Another issue that deserves investigation refers to the capacity of the autoencoder to detect binary systems where the primary is a normal star (i.e., compatible with the training set), and the secondary is too faint and contributes very little to the joint SED. That could be the case for many SB1 systems. To address this issue, we have performed a t-SNE analysis including in the input data a subset of the autoencoder training sample, i.e. a subset of normal stars. A total of 2,000 stars with known stellar parameters (temperature, gravity, and metallicity) have been considered, which represents 10\% of the training sample with know data contained in SIMBAD. The resulting t-SNE map is shown in Fig.~\ref{fig:anomaly-normal-tsne}, which shall be compared, for example, to Fig.~\ref{fig:tsne_map}. Despite the flipping along the horizontal axis, we can see that the distribution of clumps in the periphery of the map is very well preserved, and the identification of specific populations should be straightforward. Only in the central part of the map, where the normal stars are concentrated, the differences become more appreciable. In particular, there is a partial overlap between the normal stars population and part of the SB population, confirming, as expected, that many SBs would be indistinguishable from normal stars. However, there is still a large fraction of SBs that are clearly distinguished from the normal sample. These objects would deserve a detailed analysis in the future.

%%%%%%%%%%%%%%%%%%%%%
%https://sb9.astro.ulb.ac.be/mainform.cgi
%http://www.astro.gsu.edu/wds/orb6/orb6orbits.html
%https://ui.adsabs.harvard.edu/abs/2023A%26A...674A..34G/abstract
%%%%%%%%%%%%%%%%%%%%%

\section{Discussion and conclusions}\label{sect:disc_conc}

%%%%%%%%%%%%%%%%%%%%%
%Peculiar stars represent a minority compared to the total population. Therefore, in the era of large astronomical surveys, it is necessary to implement strategies for identifying objects with atypical characteristics. 
%%%%%%%%%%%%%%%%%%%%%

In this work, we demonstrate the potential of combining unsupervised machine learning methods, the photometric data from the broad-band and narrow-band filters system of S-PLUS, and stellar population information from different databases to detect potentially peculiar stars and classify stellar sub-populations. 

Although the methodology presented in this study has proven effective for identifying and classifying some stellar populations, it is worth noting that the results reported here are expected to be sensitive to the cutoff values adopted for selecting the non anomalous objects to train the autoencoder, to the threshold used to select the anomalous objects detected by the autoencoder, and to the parameters used as input for the t-SNE method. In particular, since the autoencoder has been trained using sources with magnitude $< 19$, it is probable that sources with higher magnitudes will be identified as anomalous objects even if they are not.

As for the threshold to select the anomalous objects, there is no value established in the literature, since it will depend on the objects of interest. However, a high threshold should be used if we aim to identify potentially atypical objects. In our case, we use a conservative value of 3$\sigma$ above the mean reconstruction error. This limit represents a trade-off, enabling us to obtain a sample with minimal contamination from objects classified as normal in this work, while also ensuring that we do not overlook any interesting objects. 

As for the t-SNE parameters, although the hyperparameters may influence the configuration in lower dimensions, the final result primarily depends on the perplexity value. As indicated in \citet{vanDerMaaten2008}, the choice depends on the study's objective. If the focus is on local clusters without regard to the overall configuration, small values of perplexity should be used. Conversely, if maintaining the global structure is important, large values are recommended. In our case, the perplexity was adjusted by capitalizing information from the SIMBAD database for some sources found in the over-density regions of the t-SNE map, along with visual inspection of the spectral energy distributions. While this approach may not be effective for clumpy areas without known sources in the literature, analyzing the spectral energy distributions can assist in identifying these groups.

Although a discussion on the completeness of the populations identified in this study is not the primary focus, we should stress that the size of the populations could slightly vary depending on the configuration of the input parameters in t-SNE. On the other hand, the size of the whole anomalous sample is highly dependent on the chosen threshold for the autoencoder reconstruction error. Therefore, we should expect that some sources, which could be considered anomalous, were not actually detected as such because their reconstruction error is slightly lower than the threshold.

In the end, we report 69 carbon star candidates, identified as likely CH or CEMP carbon stars based on their spatial and kinematic characteristics. We clearly identify the main sub-populations of white dwarf stars, which can be used to select objects of interest for spectroscopic follow up. Leveraging X-ray data, we demonstrate that very active low-mass stars tend to differ from the majority of their counter parts. We also detect a large number of binary systems, although it is not possible to identify specific sub-populations.

We believe that the kind of analysis presented here can serve as an alternative method for identifying candidate active stars, using multiband photometry with certain filters placed over spectral lines associated with activity. Nevertheless, we must realize that indicators derived from spectra, such as the H-alpha and Ca II H and K lines, or the analysis of epoch photometry, are undoubtedly more effective for identifying and characterizing active stars. In particular, Gaia XP data could be used to identify active stars, since the relevant lines are within its spectral range, and Gaia DR3 epoch photometry could also be applied for this purpose.

Finally, recognizing the importance of medium- and high-resolution spectroscopic follow-up to validate our findings and to better understand the nature of these anomalous objects, we are currently requesting telescope time to observe a small sample of carbon-rich star candidates reported in this work.

\clearpage
\begin{acknowledgments}
F.Q.H. acknowledges the support of a fellowship (301117/2024-1) from the PCI Program–MCTI and CNPq. F.R. acknowledges support from CNPq grant 312429/2023-1. N.H. acknowledges a fellowship 301126/2024-0 of the PCI Program - MCTI and CNPq. V.L.T. acknowledges a fellowship 302195/2024-6 of the PCI Program - MCTI  and CNPq. P.K.H. acknowledges support from the Fundação de Amparo à Pesquisa do Estado de São Paulo (FAPESP) grant 2023/14272-4. 

The authors acknowledge comments from A. Alvarez-Candal and from an anonymous referee.

The S-PLUS project, including the T80-South robotic telescope and the S-PLUS scientific survey, was founded as a partnership between the Fundação de Amparo à Pesquisa do Estado de São Paulo (FAPESP), the Observatório Nacional (ON), the Federal University of Sergipe (UFS), and the Federal University of Santa Catarina (UFSC), with important financial and practical contributions from other collaborating institutes in Brazil, Chile (Universidad de La Serena), and Spain (Centro de Estudios de Física del Cosmos de Aragón, CEFCA). We further acknowledge financial support from the São Paulo Research Foundation (FAPESP), Fundação de Amparo à Pesquisa do Estado do RS (FAPERGS), the Brazilian National Research Council (CNPq), the Coordination for the Improvement of Higher Education Personnel (CAPES), the Carlos Chagas Filho Rio de Janeiro State Research Foundation (FAPERJ), and the Brazilian Innovation Agency (FINEP). The authors who are members of the S-PLUS collaboration are grateful for the contributions from CTIO staff in helping in the construction, commissioning and maintenance of the T80-South telescope and camera. We are also indebted to Rene Laporte and INPE, as well as Keith Taylor, for their important contributions to the project. From CEFCA, we particularly would like to thank Antonio Marín-Franch for his invaluable contributions in the early phases of the project, David Cristóbal-Hornillos and his team for their help with the installation of the data reduction package jype version 0.9.9, César Íñiguez for providing 2D measurements of the filter transmissions, and all other staff members for their support with various aspects of the project.

This work used of data from the European Space Agency (ESA) mission {\it Gaia} (\url{https://www.cosmos.esa.int/gaia}), processed by the {\it Gaia} Data Processing and Analysis Consortium (DPAC, \url{https://www.cosmos.esa.int/web/gaia/dpac/consortium}). Funding for the DPAC has been provided by national institutions, in particular the institutions participating in the {\it Gaia} Multilateral Agreement.
\end{acknowledgments}

%%\appendix

\bibliography{sample631}{}
\bibliographystyle{aasjournal}

\end{document}